\documentstyle[11pt]{article}
\include{epsf}
\begin{document}

\thispagestyle{empty}

\begin{flushright}
UNIGRAZ-UTP-12-12-97 \\
hep-lat/9712015
\end{flushright}
\begin{center}
\vspace*{5mm}
{\Large On the spectrum of the Wilson-Dirac lattice operator \vskip2mm
in topologically non-trivial background configurations$^*$}
\vskip9mm
\centerline{ {\bf
Christof Gattringer}}
\vskip 2mm
\centerline{Department of Physics and Astronomy,}
\centerline{University of British Columbia, Vancouver B.C., Canada}
\vskip 5mm
\centerline{ {\bf
Ivan Hip}}
\vskip2mm
\centerline{Institut f\"{u}r Theoretische Physik} 
\centerline{Universit\"at Graz, A-8010 Graz, Austria}
\vskip10mm
\begin{abstract}
We study characteristic features of the eigenvalues of the Wilson-Dirac 
operator in topologically non-trivial gauge field configurations by 
examining complete spectra of the fermion matrix. In particular
we discuss the role of eigenvectors with real eigenvalues as the lattice 
equivalents of the continuum zero-modes. We demonstrate, that those 
properties of the spectrum which correspond to non-trivial topology 
are stable under adding fluctuations to the gauge fields. 
The behavior of the spectrum in a
fully quantized theory is discussed using QED$_2$ as an example.
\end{abstract}
\end{center}
\vskip8mm \nopagebreak \begin{flushleft} \rule{2 in}{0.03cm}
\\ {\footnotesize \ 
${}^*$ Supported by Fonds zur F\"orderung der Wissenschaftlichen 
Forschung in \"Osterreich, Projects P11502-PHY and J01185-PHY.}
\end{flushleft}
\newpage
\setcounter{page}{1}
\section{Introduction}
The Atiyah-Singer index theorem \cite{AtSi71} is considered to be among 
the greatest achievements of differential geometry. It has entered the 
physics literature in many places. Of particular interest is its 
application in gauge theories with fermions. The Atiyah-Singer index 
theorem relates the topological charge $\nu[A]$ of a sufficiently smooth 
gauge field configuration $A$ to the index of the Dirac operator on 
some compact manifold (or for gauge fields with suitable 
boundary conditions). The index of the operator can be expressed
as the difference of the number of (lineraly independent) 
negative ($n_-$) and positive ($n_+$)
chirality zero modes. The
zero modes are eigenstates of the Dirac operator
with eigenvalue $E = 0$. Since $\gamma_5$ anti-commutes 
with $i \rlap{D}{\not} \;\;$, the zero modes can be chosen as eigenstates 
$\psi_+ , \psi_-$ of $\gamma_5$
with $\gamma_5 \psi_+ = + \psi_+$ and $\gamma_5 \psi_-  = - \psi_-$.
$n_+$ ($n_-$) denotes the number of independent states
$\psi_+$ ($\psi_-$). The Atiyah-Singer index theorem \cite{AtSi71} then 
reads
\begin{equation}
\nu[A] \; = \; n_- \; - \;  n_+ \; \; \; \; .
\label{asit}
\end{equation} 
This relation allows to relate properties of fermionic observables to the
topology of the background gauge field.
Unfortunately, the gauge fields carrying the measure of the continuum path
integral do not obey the necessary smooth\-nes condition \cite{CoLa73}
for the index theorem to apply in the fully 
quantized theory. The application of the index theorem is thus restricted 
to semiclassical arguments. 

On the lattice the situation is different. It is possible to assign a
topological charge to all gauge field configurations
(see e.g.~\cite{DiGi97} for a recent review), up to so-called 
exceptional configurations which do not contribute in the continuum limit. 
The lattice path integral for the gauge fields thus 
effectively can be decomposed into topological sectors. Certainly it would 
be valuable to be able to understand the behavior of fermionic 
observables in each sector. Unfortunately there is no analytic result
for an index theorem on the lattice. However, many contributions
can be found in the literature \cite{KaSeSt86} - \cite{JaLiSiSm96}, 
where a probabilistic manifestation of the Atiyah-Singer index theorem
on the lattice and its consequences are discussed.
 
The pioneering papers \cite{KaSeSt86,SmVi87a,SmVi87} investigated
the manifestation of the index theorem on the lattice 
by analyzing the pseudoscalar density operator which through 
chiral Ward identities can be related to the topological charge. 
In two interesting papers by Smit and Vink 
\cite{SmVi87b} these relations were extended to a lattice version of the 
Witten-Veneziano formula for the pseudoscalar masses. In two numerical
studies \cite{SmVi88} the interplay between pseudoscalar density 
operators and the topological charge was further investigated for
two-dimensional models. Complete spectra of the Wilson-Dirac operator for 
non-abelian gauge fields first appeared in \cite{SeDaBa88}.

An important step in understanding the manifestation of the index theorem 
on the lattice was to realize \cite{SmVi87a,SmVi87,ItIwYo87} that the 
eigenvectors of the lattice Dirac 
operator with {\it real} eigenvalues should be interpreted as the lattice 
counterparts of the continuum zero modes. This result can best be 
understood \cite{ItIwYo87}
by using the fact that for the Wilson-Dirac operator only eigenvectors 
$\psi$ with real eigenvalues can have non-vanishing pseudoscalar matrix 
elements 
$(\psi, \gamma_5 \psi)$, similar to the zero modes in the continuum.
It was furthermore realized \cite{SmVi87a,ItIwYo87}, that only real 
eigenvalues of $D$ can give rise to exact zero eigenvalues for properly 
chosen bare mass. 
As is explicitly shown in 
\cite{ItIwYo87} this can be used to compute the topological charge by 
counting the level 
crossings of the eigenvalues of the numerically simpler hermitian 
operator $\gamma_5 D$ as
the bare quark mass is varied.
This method was later reassessed in the framework of the overlap
formalism \cite{NaNe95}. The procedure was successfully applied to study 
the interplay between topological
charge and the spectrum of $\gamma_5 D$ in SU($N$) gauge theories in 
various settings including 2d models \cite{NaNe95b} as well as
static and thermalized configurations for SU(2) and SU(3) 
\cite{NaVr97,Naetal97}. In the last reference in \cite{NaNe95} 
the index theorem is tested using the overlap framework for 
smooth abelian subgroup configurations of SU(2), for lattice 
discretizations of instantons and for a cooled configuration.

After the interest in the subject had been revived by the work of the 
overlap collaborators
\cite{NaNe95,NaNe95b,NaNeVr95,NaVr97,Naetal97} the 
last two years saw a wealth of of new articles. 
Low lying eigenvalues as well as complete spectra of the staggered 
fermion matrix were computed in \cite{Ka95,Kal95} and their distribution 
was successfully compared to the prediction of random matrix theory 
\cite{HaKaVe96}. In \cite{BaDuEiTh97} it was attempted to resolve the 
problem of exceptional configurations in quenched simulations by shifting 
the poles of the propagator due to zero modes. Studies of the behavior of 
fermionic observables in topologically non-trivial gauge field 
configurations can be found in \cite{Ne97,KoLaSi97,VeKi97}. Results for 
the spectrum in classical configurations were 
reported in \cite{GaGoMoPe97,Ne97} and in \cite{NaNeVr95,GaHiLa97b}
studies in QED$_2$ with dynamical fermions were performed. A detailed 
analysis of the lattice zero 
modes and their localization properties for the modified operators 
$\gamma_5 D$ is given in 
\cite{JaLiSiSm96}. 
\\

In this article we directly investigate properties of the 
{\it unmodified} 
Wilson-Dirac operator. In particular we compute complete spectra of 
the original Wilson-Dirac operator in sets of background configurations 
with 
known topological charge. The features of the spectrum which are due to 
the topology of the background field are analyzed. We show the 
stability of 
these features under adding fluctuations to the gauge fields thus 
demonstrating
their topological nature. The behavior of the spectrum in a fully 
quantized
theory is discussed using QED$_2$ as an example.  

The paper is organized as follows: In Section 2 we collect 
some known symmetry properties of the Wilson-Dirac operator and review 
the special role of the real eigenvalues.
This is followed by Section 3 where we con\-struct 
simple gauge field configurations to be used for numerically testing
the lattice index theorem (Section 3.1) and discuss the 
numerics (Section 3.2).
In Section 4.1 we present our numerical results for smooth gauge field
configurations, Section 4.2 contains the results for rough
configurations and in Section 4.3 we perform a stability analysis 
of the topological features of the spectrum.
Section 5 discusses the role of the index theorem in a fully quantized 
lattice gauge theory using QED$_2$ as an example. The article closes
with a discussion of the results and further perspectives (Section 6). 
 
\section{Interpreting the spectrum of the Wilson-Dirac operator}
We work on a $D$-dimensional ($D=2,4$) lattice $\Lambda$ with volume 
$L^{D-1} \times T$. Lattice sites are denoted
as $x = (x_1, ... x_D)$. The lattice spacing is set
equal to 1. We write the fermion matrix $M$ as
$M \; = \; 1 - \kappa Q$ with hopping parameter
$\kappa = (2m + 2D)^{-1}$, where $m$ is
the bare quark mass.
Due to this simple relation between $M$ and $Q$, 
in the following we will concentrate
on the analysis of the hopping matrix $Q$. $\kappa$ is the hopping 
parameter,
and the hopping matrix $Q$ is defined as (spinor and color indices 
suppressed) 
\begin{eqnarray}
Q (x,y)  &=&  \sum_{\mu = 1}^{D} Q_\mu(x,y)\;,\nonumber\\
Q_\mu(x,y)&=&
(1+\gamma_\mu) U_\mu (x - \hat{\mu}) \Delta_{x-\hat{\mu},y} \; + \;
(1 - \gamma_\mu) U_\mu (x)^\dagger \Delta_{x+\hat{\mu},y}\; .
\label{Qferm} 
\end{eqnarray}
For the $\gamma_\mu$-matrices we use the chiral representation
(see e.g.~\cite{MoMu94}). The matrix anti-commuting with $\gamma_\mu, \;
\mu = 1, ... D$ is denoted as $\gamma_5$ for both $D=2,4$. 
In order to ensure reflection positivity,
the fermions obey mixed boundary conditions, i.e.~periodic in 
$\hat{i}$-direction, $i = 1, ... D-1$, and 
anti-periodic in $\hat{D}$-direction. 
This is taken into account by the mixed 
periodic Kronecker delta $\Delta_{x\pm\hat{\mu},y}$ which is the usual
$D$-dimensional Kronecker delta, but with an extra minus sign for 
the links from $x_D = L$ to $x_D = 1$. 
The gauge fields $U_\mu(x)$
are group elements (U(1) for $D=2$ and SU($N$) for $D=4$) assigned to 
the 
links between nearest neighbors $x,x+\hat{\mu}$. They obey periodic 
boundary conditions.

It is well known, that the hopping matrix $Q$ is similar to its 
hermitian 
adjoint $Q^\dagger$ (note that $Q$ is neither hermitian nor 
anti-hermitian), 
and for even $L$ and $T$ also to $-Q$
\begin{equation}
\Gamma_5 Q \Gamma_5 \; = \; Q^\dagger \; \; , \; \; 
\Xi Q \Xi \; = \; -Q \; .
\label{simtra}
\end{equation}
The matrices $\Gamma_5$ and $\Xi$ 
implementing the similarity transformations
(\ref{simtra}) are given by $\Gamma_5(x,y) = \delta_{x,y} \gamma_5$,
$\Xi(x,y) = (-1)^{x_1 + ... x_D} \delta_{x,y}$
Both these matrices are unitary as well as hermitian and commute with each
other. We stress that for odd $L$ or $T$ the second similarity
transformation in (\ref{simtra}) does not hold since the crucial condition 
$\Delta_{x \pm \hat{\mu}, y}
(-1)^{x_1+ ... x_D+y_1+ ...y_D} = - \Delta_{x \pm \hat{\mu}, y}$
is violated for $x_\mu = 1$ (or $L,T$). In order to be able to use 
the full symmetry (\ref{simtra}) we work with even $L$ and $T$ in the
following. The overall size of the matrix $Q$ is $r \times r$ where
$r = L^{D-1} \times T \times D \times N$, with $N=1$ for gauge 
group U(1) (in $D = 2$ only). 

Beside the symmetry transformations (\ref{simtra}), $Q$ also obeys a
bound for its norm. With (\ref{Qferm}), 
using the fact that $1 + \gamma_\mu$ and $1 - \gamma_\mu$
are proportional to orthogonal projectors
and taking into account the unitarity of the $U_\mu(x)$ it is 
straightforward to show \cite{WeCh79},
$\parallel Q_\mu \parallel_\infty = \sup_{||g|| =1} 
\parallel Q_\mu g \parallel = 2$.
The norm $\parallel .. \parallel$ is defined to be the $l^2$ norm 
obtained by summing over all lattice, spinor and color indices and $g$ 
is some 
(spinor) test function on the lattice. Using the triangle inequality 
one ends up with 
$\parallel Q \parallel_\infty \; \leq \; 2D $.
This bound implies for an eigenvalue $\alpha$ of the hopping matrix $Q$
(the corresponding normalized eigenvector is denoted as $v$)
$| \alpha | = \parallel Q v \parallel \leq 
\parallel Q \parallel_\infty =  2D$.
Thus the eigenvalues of $Q$ are confined within a circle of radius
$2D$ around the origin in the complex plane. 

It follows from the symmetries (\ref{simtra}) that the
spectrum of $Q$ is symmetric with respect to reflection at both, the real 
and the imaginary axis. The eigenvalues of $Q$ 
(and due to $M = 1-\kappa Q$
also the eigenvalues of $M$) come in complex quadruplets
or in real pairs. 
\\

Let us now review the special role of the real eigenvalues and their
eigenvectors: Using the implementation of hermitian conjugation of $Q$
as a similarity transformation (\ref{simtra})
one can show \cite{ItIwYo87} the following properties of
the pseudoscalar matrix elements of eigenvectors $v_\alpha, v_\beta$ 
of $Q$ with eigenvalues $\alpha, \beta$  
\begin{eqnarray}
{v_\alpha}^\dagger \; \Gamma_5 \; v_\alpha & \neq & 0 
\; \; \; \; \; \mbox{only for} \; \; \; \alpha \in \; 
\mbox{I\hspace{-1.1mm}R} \; ,
\nonumber \\
{v_\alpha}^\dagger \; \Gamma_5 \; v_\beta & \neq & 0 
\; \; \; \; \; \mbox{only for} \; \; \; \; \alpha \; = \; 
\overline{\beta} \; .
\label{chirlat}
\end{eqnarray}
only for {\it real} eigenvalues $\alpha $. 
\vskip3mm
\noindent
These equations should be compared to the chiral properties
of the eigenstates of the (anti-hermitian) Dirac operator 
$\not \mbox{\hspace{-1.6mm}} D$ in the continuum. 
Denoting an eigenstate with eigenvalue
$iE$ as $\psi_E$ one finds (since $\gamma_5$ anti-commutes with 
$\not \mbox{\hspace{-1.4mm}} D$)
\begin{eqnarray}
( \psi_{E} \; , \; \gamma_5 \; \psi_E ) & \neq & 0 
\; \; \; \; \; \mbox{only for} \; \; \; \; E = 0 \; , 
\nonumber \\
( \psi_{E^\prime} \; , \; \gamma_5 \; \psi_E ) & \neq & 0 
\; \; \; \; \; \mbox{only for} \; \; \; \; E^\prime = - E \; .
\label{chircont}
\end{eqnarray}
When comparing (\ref{chirlat}) and (\ref{chircont}) 
it is immediately clear,
that only the eigenvectors of $Q$ (and thus of $M$) with {\it real}
eigenvalues can be the lattice equivalents of the zero modes in the 
continuum. Based on this observation one can interpret the 
Atiyah-Singer index theorem on the lattice in the form 
\cite{ItIwYo87}
\begin{equation}
\nu[U] \; = \; R_- \; - \; R_+ \; .
\label{lasit}
\end{equation}
Here $R_+$ ($R_-$) denotes the number of {\it real} eigenvalues $\alpha$
of the hopping matrix $Q$ with positive (negative) matrix element 
${v_\alpha}^\dagger \; \Gamma_5 \; v_\alpha$ in the physical branch 
of the spectrum, i.e.~in the vicinity of +8 in the complex 
plane.

Also the real eigenvalues in the other branches of the spectrum,
corresponding to corners of the Brillouin zone different from $(0,0,0,0)$,
obey a simple pattern governed by the topological charge of the gauge
field configuration: The right hand side of (\ref{lasit}) simply has to
be multiplied by a factor corresponding to the degeneracy of the 
corresponding
corner of the Brillouin zone (e.g.~4 for the $(\pi,0,0,0)$-type corners,
6 for the $(\pi,\pi,0,0)$-type corners etc.) and an extra minus sign for 
each $\pi$ in the coordinate of the corner (compare Table 1,
Fig.~\ref{schempic} and the discussion in Section 4). 
(\ref{lasit}) is expected to hold
for configurations with plaquette elements sufficiently close to 1,
such that exceptional configurations are avoided.

We remark that when analyzing the manifestation of 
the index theorem on the lattice the existence of Eq. (\ref{chirlat}) and 
its interpretation (\ref{lasit}) makes the Wilson 
formulation more convenient compared to 
the staggered form of the fermion matrix.
The latter formulation gives rise to an anti-hermitian matrix with only
purely imaginary eigenvalues and thus lacks a natural criterion to 
identify the lattice equivalents of the zero modes. 

\section{Numerical analysis of topological features of the spectrum}
\subsection{Background configurations with non-trivial to\-po\-lo\-gy}
In order to test the form (\ref{lasit}) of the 
Atiyah-Singer index theorem on the lattice, one has to compare 
the number of the real eigenvalues to the topological charge of 
the gauge field configuration. For this enterprise it is ne\-ces\-sary to 
use a definition of the topological charge independent from the Dirac
operator, such as the geometric definition \cite{Lu82,PhSt86}.  
However, although conceptionally most satisfying,
the geometric definition 
is rather costly to evaluate for $D = 4$ with gauge group SU($N$) 
\cite{FoGiLaSch85}. 
For a first analysis of (\ref{lasit}) in $D = 4$ we use sets of 
topologically non-trivial SU(2) gauge field configurations with known 
topological charge \cite{SmVi87a}. These are configurations that
are a lattice transcription of continuum SU(2) gauge fields with constant
field strength on a torus. They are given by 
\begin{eqnarray}
U_1(x) & = & \exp \Big( -i \tau \; [
2 \pi \frac{s}{L^2} (x_2 -1) + \varphi_1 ] \Big) \; ,
\nonumber \\
U_3(x) & = & \exp \Big( -i \tau \; [
2 \pi \frac{t}{LT} (x_4 -1) + \varphi_3 ] \Big) \; ,
\nonumber \\
U_2(x) & = & \left\{ \matrix{ 1 & \; \; \mbox{for} \; x_2 \neq L \cr
\exp \Big( i \tau \; [ 2 \pi \frac{s}{L} (x_1 -1) + \varphi_2 ] \Big)
& \; \; \mbox{for} \; x_2 = L \cr }
\right. \; \; \; ,
\nonumber \\
U_4(x) & = & \left\{ \matrix{ 1 & \; \; \mbox{for} \; x_4 \neq T \cr
\exp \Big( i \tau \; [ 2 \pi \frac{t}{L} (x_3 -1) + \varphi_4 ] \Big) 
& \; \; \mbox{for} \; x_4 = T \cr }
\right. \; \; \; .
\label{smoothu}
\end{eqnarray}
$\tau$ is one (arbitrary) of the SU(2) generators (Pauli matrices). We 
remark that the spectrum of the hopping matrix $Q$ does not depend on the 
choice of $\tau$, as can be seen from a hopping expansion of 
its characteristic polynomial. 
$s, t$ are integers. We generalized the configurations given in 
\cite{SmVi87a} by introducing constant phases $\varphi_\mu$ which leave
the topological charge invariant, but allow for a wider range of simple
configurations where we can check the index theorem (\ref{lasit}).
For the configurations (\ref{smoothu}) the plaquette elements 
$U_{\mu \nu}(x) = 
U_\mu(x)^\dagger U_\nu(x + \hat{e}_\mu)^\dagger U_\mu(x + \hat{e}_\nu)
U_\nu(x)$ are constant with
$U_{1 2}(n) = \exp\big( - i 2\pi \tau s/L^2\big)$, 
$U_{3 4}(n) = \exp\big( - i 2\pi \tau t/LT\big)$ and 
all other plaquettes $U_{\mu \nu} = 1$.
The topological charge is given by 
\begin{equation}
\nu[U] \; = \; 2 s t \; .
\end{equation}
The configurations (\ref{smoothu}) have action $4 L^3 T [ 2 - 
\cos(2\pi s/L^2) - \cos(2 \pi t/ LT) ]$ which for large $L=T$
approaches the value $8 \pi^2 (s^2 + t^2) \geq 8 \pi^2 2 |st|
= 8 \pi^2 |\nu|$, thus obeying in this limit the lower bound for the 
(continuum) gauge field action in the sector $\nu$.

The fields (\ref{smoothu}) are maximally smooth (in terms of their
plaquettes). However small fluctuations around the fields 
(\ref{smoothu}) should leave the topological charge
and the properties of $Q$ which depend on it (number of real eigenvalues,
chiral properties of the eigenvectors) invariant. For large enough 
$L,T$ and small enough fluctuations the invariance of the 
(geometric) topological charge is an exact result: L\"uscher
\cite{Lu82} gives the (non-optimal)
bound Tr$[1 - U_{\mu \nu}(x)] < 0.03$ for a SU(2) configuration 
to be non-exceptional (i.e. the configuration
can be assigned a topological charge; 
in the work of Phillips and Stone \cite{PhSt86} a similar bound is given).
In order to change the 
topological sector, the gauge fields have to go through an
exceptional configuration. Thus if the fluctuations are small enough,
the topological charge remains invariant.

In order to analyze 
configurations fluctuating around (\ref{smoothu}) we define new 
link variables
\begin{equation}
U_\mu(x)_{rough} \; \; = \; \; U_\mu(x)_{(\varepsilon)} \;
U_\mu(x)_{old} \; ,
\label{roughu}
\end{equation}
where the smooth link variables $U_\mu(x)_{old}$ are given by
(\ref{smoothu}), and the fluctuations $U_\mu(x)_{(\varepsilon)}$ 
are SU(2) elements in the vicinity of 1, defined as ($\sigma_j, \;
j = 1,2,3$ denote the Pauli matrices)
\begin{equation}
U_\mu(x)_{(\varepsilon)} \; \; = \; \;
\mbox{1\hspace{-1.1mm}I}_2 \; \; r^{(0)}_\mu(x) \; \; + \; \; 
i \sum_{j=1}^3 \sigma_j  \; r^{(j)}_\mu(x) \; ,
\end{equation}
where $r^{(j)}_\mu(x), \; j = 1,2,3$ are sets of small random
numbers with
\begin{equation}
-\varepsilon \leq r^{(j)}_\mu(x) \leq \varepsilon \; \; , \; \; 
j = 1,2,3 \; \; \; \; 
\mbox{and} \; \; \; r^{(0)}_\mu(x) = 
\sqrt{1 - \sum_{j=1}^3 \Big( r^{(j)}_\mu(x) \Big)^2} \; .
\end{equation}
The size of the fluctuations is bounded by $\varepsilon$ and in 
the limit $\varepsilon \rightarrow 0$ we have $U_\mu(x)_{(\varepsilon)}
\rightarrow 1$. In the numerical analysis described below, we will 
use values
$\varepsilon = 0.15$ and 0.3 and apply the roughening procedure 
(\ref{roughu})
several times ($n_r$ times). Certainly the gauge field action increases 
with 
roughening. We found, that when applying several runs of the roughening 
procedure with different values of
$\varepsilon$, the average values of the action plotted as 
a function of $\varepsilon \sqrt{n_r}$ lie on a universal curve. 
The roughening 
procedure thus behaves like a directed random walk 
towards higher values of
the action.

In order to avoid confusion, we finally remark that the configurations 
(\ref{smoothu}), (\ref{roughu}) do not dominate the path integral
of lattice gauge theory 
(see e.g.~\cite{FoGaSt97,GrHaKo97,GaHiLa97a}). 
They are merely a convenient
tool to analyze the spectrum in gauge field configurations with known
topological charge.

\subsection{Diagonalization of $Q$ and checks} 
\noindent
Computing all eigenvalues of the fermion matrix and some of the 
eigenvectors is a challenging problem already for rather small,
four-dimensional lattices. 
The fermion matrix for SU($N$) has size $r \times r$ with 
$r = L^{D-1} \times T \times D \times N$.
Studying e.g.~SU(2) on a $4^4$ lattice means diagonalizing
and computing eigenvectors for a $2048 \times 2048$ complex
matrix. Storing the whole matrix in double precision requires
already 64 MB of memory. 
Doubling the lattice size increases $r$ by a factor 
of 16, and the required memory by a factor of 256. 
Thus the memory requirements restrict one to rather small lattices. 
In particular we studied
the spectrum for SU(2) gauge fields on lattices of size 4$^4$ and
$4^3 \times 6$. 

We remark that for hermitian matrices there exists a method with 
modest storage
requirements: the Lanczos method without reorthogonalization
\cite{CulWil85}. It was successfully used by e.g.~Kalkreuter and Simma
\cite{Kal95} to diagonalize the hermitian operator $\gamma_5 M$. 
Unfortunately, for the unchanged, non-hermitian Wilson-Dirac matrix 
$M$ (or $Q$ respectively) the situation is less suitable for the 
Lanczos method without reorthogonalization. 
Barbour et al.~\cite{SeDaBa88,BarLea92} report
that they were not able to compute all the eigenvalues and they
were forced to reorthogonalize. In that case, the Lanczos vectors
have to be stored and storage requirements become comparable with 
standard methods. Another typical problem for the Lanczos method
without reorthogonalization is that the eigenvalues come with
wrong multiplicities \cite{CulWil85,Kal95}. Since the exact
number of degenerate eigenvalues is essential for our 
investigation, the use of standard routines for general complex
matrices from the LAPACK package turned out to be the best choice: 
$Q$ was first diagonalized (using LAPACK routines ZGEHRD, ZHSEQR) 
and then 
eigenvectors for
particular eigenvalues were computed using ZHSEIN and ZUNMHR. 
Computing the spectrum
(without eigenvectors) for the $4^4$ lattice takes typically 4 hours of 
computer-time on a MIPS R10000 CPU at 180 MHz. The time
for the computation of a single eigenvector to a given eigenvalue
is only several seconds. 

Our investigation requires the distinction between real eigenvalues and
eigenvalues with non-vanishing imaginary part. For some of the gauge field
configurations we encountered complex eigenvalues with rather small
imaginary parts of order $10^{-3}$
but there was never a problem distinguishing 
them from truly real eigenvalues, which in the numerical computation have 
imaginary parts smaller than $10^{-13}$ and do not have a complex
conjugate partner. A second criterion to distinguish 
the eigenvalues is of course the pseudoscalar matrix element of the
corresponding eigenvectors, which is of order 1 for real eigenvalues, 
but of order $10^{-15}$ for eigenvalues with 
non-vanishing imaginary part. 

There are several ways to test the correctness and accuracy of the 
programs. A first check we performed is comparing the 
numerical result for the trivial case (all link variables = 1) 
to the analytic result for the spectrum obtained by Fourier transformation. 
Our diagonalization routine reproduces the analytic result 
with an accuracy of $10^{-12}$. 
 
A more advanced test makes use of results known from the 
hopping expansion (see e.g.~\cite{MoMu94}). 
Using the fact, that $1 \pm \gamma_\mu$
are proportional to projectors, one can show, that Tr$Q^n$
is proportional to the sum over traces over products of links 
ordered along non-backtracking, closed paths of length $n$. 
In particular
\begin{equation}
\mbox{Tr} \; Q^2 \; = \; \sum_{i=1}^{r} {\alpha_i}^2 \; = \; 
0 \; \; , \; \; 
\mbox{Tr} \; Q^4 \; = \;\sum_{i=1}^{r} {\alpha_i}^4 \; = \; 
-16 D \sum_p \mbox{Re Tr} \; U_p \; .
\label{sumrule}
\end{equation}
When some of the sides of the lattice $L$ or $T$ have length 4,
Tr$Q^4$ has some extra terms due to loops around the boundary.
Sums over odd powers of eigenvalues $\alpha_i$ 
vanish trivially, since eigenvalues come always in pairs $\alpha$
and $-\alpha$ (compare (\ref{simtra})).
We checked the quadratic and quartic sum rules, and found that they are 
obeyed with an accuracy of $10^{-9}$ for all spectra we analyzed.

We remark, that from the hopping expansion it is 
also obvious that the spectrum of the Wilson-Dirac 
operator is gauge invariant: The characteristic polynomial
$\det( \alpha - Q )$ has a hopping expansion in terms of Tr$Q^n$.
Each of these contributions is a sum over traces of 
closed loops of link variables
and thus gauge invariant. This implies gauge invariance of the 
characteristic
polynomial and thus of the spectrum. An alternative argument 
for the gauge invariance of the spectrum is to use the fact that the 
Wilson-Dirac 
operator transforms in the adjoint representation of the gauge group,
and thus the eigenvalues remain invariant (the eigenvectors transform 
in the fundamental representation).  

\section{Numerical results}
\subsection{Smooth configurations}
\begin{figure}[htpb]
\epsfysize=4.8in
\hspace*{-4mm}
\epsfbox[109 418 510 774] {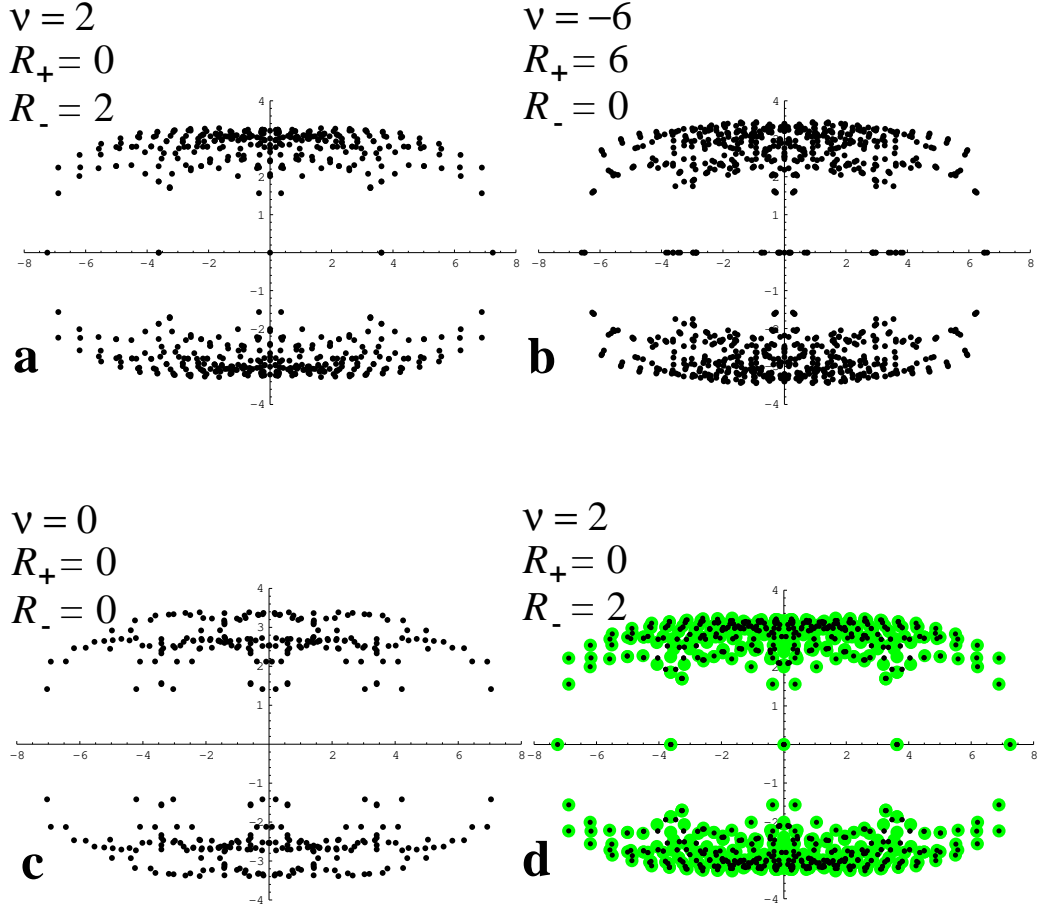}
\caption{Plots of the spectra of $Q$ in the complex plane,
for smooth background configurations
(\protect{\ref{smoothu}}) with $s = 1, t = 1$ (a), 
$s = 1, t = -3$ (b), $s = 1, t = 0$ (c) and
for $s = 1, t = 1$ (d). For Plots a, b, c all phases $\varphi_\mu$ 
were chosen to be zero,
while for Plot d we have $\varphi_\mu = 2\pi/7, \mu = 1,2,3,4$ 
(small black dots)
and compare it with the spectrum for the zero phase 
configuration (larger grey dots). $\nu$ is the 
topological charge of the gauge field used ($= 2st$) 
and $R_+, R_-$ are the numbers of
real eigenvalues in the physical branch of the spectrum
with positive and negative chirality.
\label{smoothspect}}
\end{figure}

We start presenting our numerical results with the discussion of the
spectrum for the smooth fields (\ref{smoothu}). We analyzed more than 
50 of these configurations with different values 
of the parameters\footnote{
As remarked above, it can be shown that the spectrum of the 
Wilson-Dirac operator is 
independent of $\tau$, but different choices of $\tau$ were
used to check the program.} 
$\tau, s, t, \varphi_\mu$ for lattices with $L = 4$ and $T = 4, 6$.
Fig.~\ref{smoothspect} shows 4 spectra of $Q$ on a $4^4$ lattice
displaying different
features of the spectrum which we will discuss below. 

Fig.~\ref{smoothspect}.a shows the spectrum for the $s = 1, t = 1$
configuration, which has topological charge $\nu$ = 2. 
The first thing to note are the expected symmetries of the spectrum with 
respect to reflection at real and imaginary axis. All eigenvalues
lie within a circle of radius 8 around the origin in the complex plane, 
as
discussed in Section 2 and their imaginary parts are bounded by 4. 
The complex eigenvalues are well separated from the real axis, and many 
of them are degenerate (note that there are all together 2048 
eigenvalues).

\begin{figure}[htp]
\epsfysize=0.75in
\hspace*{20mm}
\epsfbox[28 639 332 704] {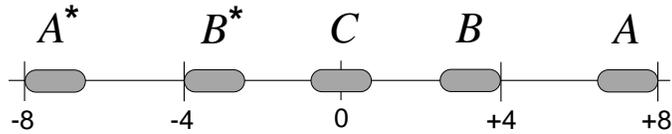}
\caption{Schematic picture of the branches containing the real part of 
the spectrum of $Q$.  
\label{schempic}}
\end{figure}

There are 32 real eigenvalues, all of them multiply degenerate. 
They lie in the vicinity of 
the values -8, -4, 0, 4, 8. We denote these branches where the real 
eigenvalues
are found as $A^*, B^*, C, B, A$ as depicted in Fig.~\ref{schempic} 
and
described in Table 1. As discussed in Section 2
Theorem, these branches correspond to the poles of the free 
propagator with periodic boundary conditions, situated at the corners
of the Brillouin zone:
$(0,0,0,0), (\pi,0,0,0) + 3$ permutations, etc. For the $s = 1, t = 1$  
configuration we find that the eigenvalues in the branches $A^*, B^*, 
C, B, A$
are degenerate and the branches are occupied by 2, 8, 12, 8 and 2 real 
eigenvalues. The pseudoscalar matrix
elements $v^\dagger \; \Gamma_5 v$ have the same sign for all eigenvalues 
within a branch, with the signs distributed among the branches as
$-1, +1, -1, +1, -1$. All these features
are as described in the lattice version (\ref{lasit}) of the index theorem
on the lattice and the right hand side terms of (\ref{lasit}) for the
principal branch are given by $R_+ = 0, R_- = 2$.

Fig.~\ref{smoothspect}.b and c show spectra for configurations with
$s = 1, t = -3$, having topological charge $\nu = -6$, 
and $s = 1$ and $t = 0$, with topological charge $\nu = 0$. For the 
former case we find that the numbers of real eigenvalues in each branch
have increased to 6, 24, 36, 24, 6. The signs of the pseudoscalar matrix
elements are again equal for each branch, but now given by 
+1, -1, +1, -1, +1, since the topological charge is negative for this 
configuration. Thus $R_+ = 6, R_- = 0$.
For the spectrum in Fig.~\ref{smoothspect}.c, which is for 
a configuration with zero topological charge, we find no real 
eigenvalue and
$R_+ = R_- = 0$, compatible with (\ref{lasit}).

Finally Fig.~\ref{smoothspect}.d shows the spectrum for the s = 1, t = 1
case but now for all link variables we have chosen the non-zero value 
$\varphi_\mu = 2\pi/7, \mu = 1,2,3,4$ for the phase in (\ref{smoothu}) 
(small
black dots in Fig.~\ref{smoothspect}.d). The large grey dots indicate the 
spectrum for the configuration with zero phase. The figure nicely
demonstrates, that the phase can shift several of the complex eigenvalues, 
but the real part of the spectrum, 
which reflects the topological charge (unchanged by the phase)
remains invariant.

The general picture which emerges after our analysis of various spectra for
smooth configurations (\ref{smoothu}) is summarized in Fig.~\ref{schempic} 
and Table 1. It entirely confirms formula (\ref{lasit}).

\begin{table}[htp]
\begin{center}
\begin{tabular}{|l|c|c|c|c|c|}
\hline
branch  & $A^*$ & $B^*$ & $C$ & $B$ & $A$ \\
\hline 
$R_- - R_+$ & $\nu$ & $-4\nu$ & $6\nu$ & $-4\nu$ & $\nu$ \\
\hline
corners & $(\pi,\pi,\pi,\pi)$ & $(\pi,\pi,\pi,0)$ 
& $(\pi,\pi,0,0)$ & $(\pi,0,0,0)$ & $(0,0,0,0)$ \\
of B.Z. & & + 3 perm. & + 5 perm. & + 3 perm. &  \\
\hline
\end{tabular}
\end{center}
\caption{Properties of the branches containing the real part 
of the spectrum of $Q$.}
\end{table}

\subsection{Rough configurations}

After testing the lattice version (\ref{lasit}) of the index theorem  
for the smooth fields (\ref{smoothu}), we consider now the roughened
fields (\ref{roughu}). One expects, that for a small amount of roughening, 
which does not destroy the topological charge of the gauge field 
configuration, the relation (\ref{lasit}) remains invariant.

\begin{figure}[htpb]
\epsfysize=4.35in
\hspace*{-4mm}
\epsfbox[32 313 434 641] {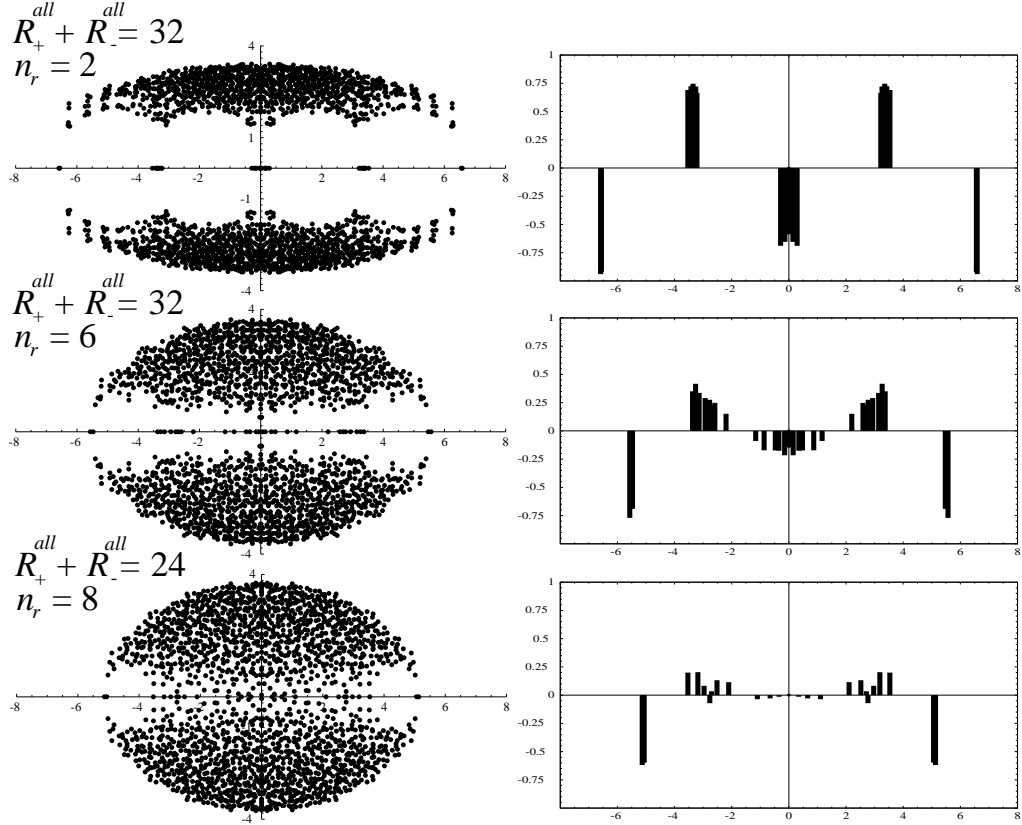}
\caption{Behavior of the spectrum of $Q$ (left-hand side plots) and the
pseudoscalar matrix elements $v^\dagger \Gamma_5 v$ plotted
at the positions of the corresponding real eigenvalues
(right-hand side plots) for the
$s = 1, t = 1$ configuration when the roughening procedure
(\protect{\ref{roughu}}) is applied. We show the results for roughening 
steps
$n_r = 2, 6$ and 8, each with $\varepsilon = 0.3$. 
For $n_r = 2,6$ (upper four plots) the total number of real eigenvalues 
in the spectrum is $R_+^{all} + R_-^{all}$ = 32, 
according to (\protect{\ref{lasit}}), while for 
$n_r = 8$ the gauge field becomes too rough, and (\protect{\ref{lasit}})
becomes violated giving rise to only 24 real eigenvalues in the spectrum.
\label{roughspect}}
\end{figure}

In Fig.~\ref{roughspect} we show the spectra of $Q$ (left-hand side 
plots) and the pseudoscalar matrix elements $v^\dagger \Gamma_5 v$ 
plotted at the positions of the corresponding real eigenvalues
(right-hand side plots) for the
$s = 1, t = 1$ configuration having $\nu = 2$. 
We show the results for roughening steps
$n_r = 2, 6$ and 8, each step performed with $\varepsilon = 0.3$
(see (\ref{roughu})).
Comparing the
spectra for the roughened configurations with the corresponding smooth 
spectrum Fig.~\ref{smoothspect}.a, one finds that the degeneracy of the 
eigenvalues is lifted. For $n_r = 2,6$ the real eigenvalues are 
still more or less
concentrated in the branches $A^*, B^*, C, B, A$, but this structure 
becomes
less pronounced with increasing number of roughening steps. 
For $n_r = 2, 6$
the total number of real eigenvalues is $R_+^{all} + R_-^{all} = 32$ 
as expected from the index theorem 
($= 16 \times |\nu|$). For $n_r = 8$ the separation of the real spectrum 
into
distinct branches has almost vanished, and the total number of real 
eigenvalues 
has dropped to $R_+^{all} + R_-^{all} = 24$. 
Obviously the configuration becomes 
too rough and (\ref{lasit}) does no longer hold. It is also interesting 
to note, that the complex part of the spectrum approaches the real line.
This trend continues when $n_r$ is increased further, and the distribution 
of
the eigenvalues becomes uniform inside the disk of radius 4 around the 
origin for random configurations (compare \cite{SeDaBa88}). 
This observation
nicely fits into the picture obtained from a strong coupling expansion.
It can be shown, that the divergence of the pion propagator in strong
coupling \cite{Wi75} is equivalent to a bound of 4 for the modulus
of the eigenvalues of $Q$.

The plots for the pseudoscalar matrix elements on the right hand side of 
Fig.~\ref{roughspect} confirm the obtained picture. For $n_r = 2,6$ the 
matrix elements $v^\dagger Q v$ all have the same sign within the branches
of the real spectrum. For $n_r = 8$ some of the real eigenvalues, and thus 
their matrix elements have vanished, and the remaining matrix elements 
do no longer obey the sign pattern of the smoother configurations. 

\begin{figure}[htpb]
\epsfysize=2.1in
\hspace*{-1mm}
\epsfbox[72 440 476 609] {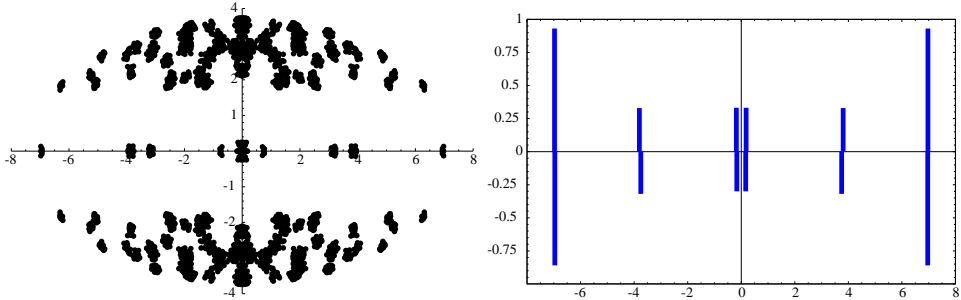}
\caption{Spectrum of $Q$ (left-hand side plot) and the corresponding
matrix elements $v^\dagger \Gamma_5 v$ (right-hand side plot)
for the $s = 0, t = -2$ configuration after one step of roughening
with $\varepsilon = 0.3$. 
The gauge field has $\nu = 0$ and according to formula 
(\protect{\ref{lasit}}), the sum of the signs of the matrix elements has 
to 
cancel for each branch of the spectrum. This feature is obvious from the 
right-hand side plot.
\label{nu0spect}}
\end{figure}

So far we presented only configurations where the matrix elements
$v^\dagger \Gamma_5 v$ have the same sign for all real eigenvalues in
a branch of the spectrum. We found, that our test configurations 
(\ref{smoothu}), (\ref{roughu}) also allow for cases where both signs 
occur within a branch. Due to the anti-periodic boundary conditions in 
time direction, configurations of the type
$s = 0, t \neq 0$, although they have $\nu = 0$,
give rise to spectra with an even number of real eigenvalues in each 
branch of the spectrum. Half of the corresponding eigenvectors
have positive and half of them 
negative chirality and their contributions to the right hand
side of (\ref{lasit}) cancel. Fig.~\ref{nu0spect} shows the
spectrum for the $s = 0, t = -2$ configuration after 1 roughening step
(left-hand side plot) and the corresponding pseudoscalar matrix elements 
$v^\dagger \Gamma_5 v$ in all branches of the spectrum (right-hand 
side plot). 
Although the 
configuration has $\nu = 0$ there are 12 real eigenvalues in the spectrum.
The index theorem on the lattice (\ref{lasit})
is nevertheless fulfilled, since the
signs of the pseudoscalar matrix elements cancel in each branch of the 
spectrum.

The left-hand side plot in Figure 
\ref{nu0spect} also demonstrates nicely another important result of
our analysis: The modulus of eigenvalues is no proper criterion to 
identify
zero modes. From the spectrum plot in Fig.~\ref{nu0spect} it is obvious, 
that in the close vicinity of the
real eigenvalues there cluster several complex eigenvalues. These complex
eigenvalues have nothing to do with zero modes, since their pseudoscalar
matrix elements vanish identically ($\sim 10^{-15}$ in our numerical 
analysis),
while the matrix elements for the real eigenvalues are of order 1.

We remark, that increasing the size of one of the lattice sides to 6
does not affect the structure of the spectrum. For all sufficiently
smooth configurations we analyzed, the lattice version (\ref{lasit}) 
of the 
index theorem was confirmed, clearly demonstrating that the structure 
of the real eigenvalues is the part of the spectrum which remains 
invariant 
within the topological sectors.

\subsection{Stability analysis for the topological features of the 
spectrum}
In this section we present results from the analysis of the spectra for
a whole sample of roughened gauge fields. This serves to demonstrate that
the topological features of the spectrum are stable for a set of 
rough configurations (\ref{roughu}) fluctuating around the smooth 
configurations (\ref{smoothu}).

\begin{figure}[htb]
\epsfysize=2.7in
\hspace*{10mm}
\epsfbox[62 448 525 754] {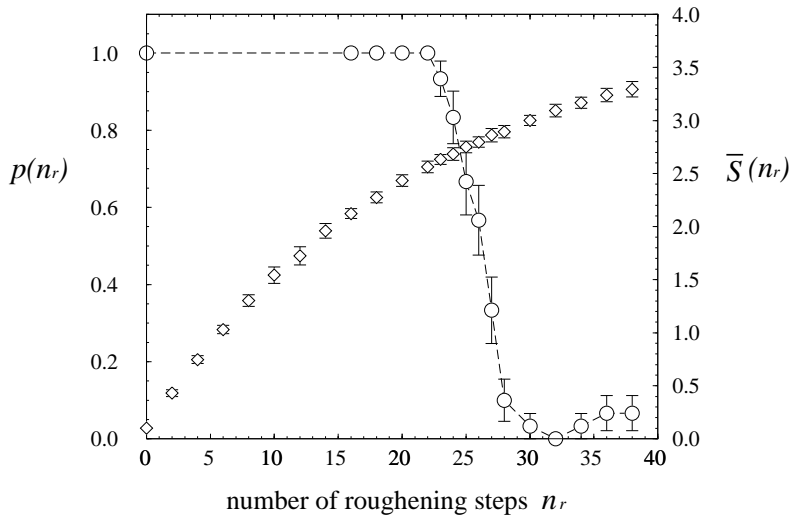}
\caption{Probability $p$ of finding (\protect{\ref{lasit}}) 
correct (circles, left vertical scale) as a function 
of the number of roughening steps $n_r$, each with $\varepsilon = 0.15$. 
We also show the average value
of the gauge field action per plaquette $\overline{S}$ 
(diamonds, right vertical scale).
\label{stabiplot}}
\end{figure}

In particular we analyzed the spectra for roughened gauge field 
configurations
starting from the smooth configuration with $s=t=1 (
\nu = 2)$. We then applied 
$n_r$ steps of the roughening procedure each with $\varepsilon = 0.15$.
For all together 15 values of $n_r$ in the range of $n_r = 16$ to 
$n_r = 38$
we computed spectra for 30 configurations at each $n_r$. For each sample
of 30 spectra (at fixed $n_r$) we analyzed the topological features of
the eigenvalues: The smooth configuration with $t=s=1$ has all together
32 real eigenvalues, distributed among the branches as $2, 8, 12, 8, 2$.
The configuration has $\nu = 2$, and thus according to (\ref{lasit})
the quoted numbers are the
{\it minimal} amount of real eigenvalues in each branch. In order to
be compatible with (\ref{lasit}) there is 
always the possibility that there are more real eigenvalues in a branch,
but with different chirality of the corresponding eigenvectors, such 
that the
difference of left- and right-handed contributions gives the numbers
2,8,12,8,2
for the branches. However, any spectrum with less than 32 real eigenvalues
clearly violates (\ref{lasit}). In almost all cases we analyzed, the 
breakdown
of (\ref{lasit}) was signalized by such a drop of the number of real 
eigenvalues 
below 32. For the following discussion we adopt as our criterion 
for a breakdown of (\ref{lasit}) a violation of
\begin{equation}
\# \; of \; real \; eigenvalues \; \; = \; \; 32 \; .
\label{viocrit}
\end{equation}
We define the probability function $p(n_r)$ to be the probability
of finding (\ref{viocrit}) correct, and compute it for each sample of 
30 spectra at fixed $n_r$. 
Fig.~\ref{stabiplot} shows the probability $p(n_r)$ (circles, scale on 
the 
left vertical axis) for different 
values of $n_r$. In the figure we also display the average value of 
the gauge field action per plaquette $\overline{S}$ 
(diamonds, right vertical scale).

The figure shows nicely, that there is a rather large number of roughening 
steps where the topological structure of the spectrum remains unperturbed,
i.e. $p = 1$ (from $n_r = 0$ to $n_r = 22$). 
During these roughening steps the action per plaquette $\overline{S}$
increases (starting with a value of 0.1 at $n_r = 0$)
roughly by a factor of 25 which demonstrates the stability of the 
topological structure in more physical units. Then after 22 roughening 
steps
the probability $p$ for the index theorem (\ref{lasit})
to hold drops sharply reaching a level 
of $p \sim 0.05$ at $n_r = 30$. Under further roughening the
probability function keeps fluctuating around this value. The plot nicely
demonstrates the stability of the topological structure of the spectrum.
In particular real eigenvalues and the pseudoscalar matrix elements of their 
eigenvectors are seen to be rather stable which underlines their topological 
nature.

We performed the same analysis (with less statistics) for roughened
gauge fields starting from the smooth $t = s = 2$ configuration.
Again we observed a large region where the topological structure of the 
spectrum remained stable, with a rather sharp drop of the probability 
function at the end of the region. 
We found that this drop was at a different value
of the action per plaquette (approximately 25 \% smaller), indicating that,
at least for the lattice size we considered, there is no universal threshold
in terms of the action where a topological sector gets destroyed.

\section{Discussion of the overall picture in the fully quantized
theory using QED$_2$ as an example} 
In the last section we have numerically investigated the  
form (\ref{lasit}) of the index theorem on the lattice using test 
configurations for 4-D SU(2) lattice gauge theory.
The set of gauge field configurations 
(\ref{smoothu}), (\ref{roughu}) we used is of course rather restricted. 
On the other hand, testing the 
obtained picture in a simulation of QCD$_4$ is certainly too demanding
for present computer technology. However it is possible 
\cite{GaHiLa97b} to analyze the
Atiyah-Singer index theorem (\ref{lasit}) in the simpler model of QED$_2$. 

QED$_2$ is not only considerably cheaper to simulate, but also has a 
very simple expression \cite{toplat2} for its topological charge:
$\nu[U] \; = \; \frac{1}{2\pi} \sum_x \theta_p(x)$,
where $\theta_p(x) = -i \ln U_{12}(x)$ with $\theta_p(x) \in (-\pi, \pi)$.
Configurations where $U_{12}(x) = -1$ for one of the plaquettes are 
so-called
exceptional configurations \cite{Lu82}, and no topological charge can
be assigned to them. They are of measure zero in the path integral.

\begin{figure}[htb]
\epsfysize=2.7in
\hspace*{19mm}
\epsfbox[12 166 390 465] {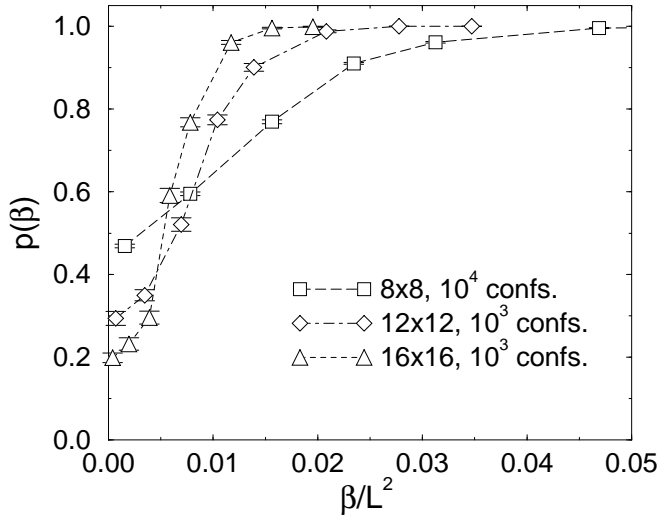}
\caption{Probability $p$ of finding (\protect{\ref{nre}}) 
correct as a function 
of $\beta/L^2$. We show our results for lattice sizes $L=8, 12$ and 16. 
The symbols are connected to guide the eye. 
\label{qed2plot}}
\end{figure}

Again it is possible to use the Atiyah-Singer index theorem 
in the continuum as a guiding line
of what to expect on the lattice. For QED$_2$ there holds another index
theorem, sometimes referred to as {\it Vanishing Theorem} \cite{top2}.
It states that there are either only left- or only right-handed zero modes.
Furthermore, taking into account 
that in $D=2$ there are 4 species of fermions, 1
in the physical branch and 3 doublers, one expects to find
\begin{equation}
\# \; of \; real \; eigenvalues \; = \; 4 \Big|\nu[U] \Big| \; .
\label{nre}
\end{equation}
We define $p$ to be the probability of finding (\ref{nre}) correct
and evaluate this probability for a whole ensemble of Monte Carlo 
configurations generated in a simulation of QED$_2$ with dynamical 
fermions \cite{GaHiLa97b}. We computed $p$ for several values of $L = T$
and $\beta$. The hopping parameter $\kappa$ was always chosen close to 
the critical value for the given $L$ and $\beta$. In Fig.~\ref{qed2plot}
we plot\footnote{We thank A.~van der Sijs for a discussion concerning the
presentation of this data.} the 
probability function $p$ as a function of the 
dimensionless ratio $\beta/L^2$.

It is obvious from this plot that when sending
$L \rightarrow \infty$ and holding $\beta/L^2$ fixed   
(which corresponds to sending $L \rightarrow \infty$
while holding $\xi/L$ fixed), the probability
function $p$ approaches 1. This implies that (\ref{nre}) 
holds with probability 1 in the continuum limit. A detailed
analysis \cite{GaHiLa97b} shows, that this result is essentially
uniform in $\nu$ and also the chiral properties of the 
eigenvectors are conform with the expectations from Atiyah-Singer Index
Theorem and Vanishing Theorem.

It has to be stressed, that 
this result is not restricted to some subset of configurations 
(as is our analysis for QCD$_4$) since it was established using 
the complete set of Monte-Carlo configurations.  This demonstrates
that the lattice version of the index theorem holds in fully
quantized QED$_2$. We remark, that in \cite{NaNeVr95} the interplay 
between the topological charge and the spectrum of the lattice Dirac 
operator was further investigated by demonstrating within the overlap 
formalism that the chiral condensate in the Schwinger model 
originates from the sectors with topological charge $\pm 1$.

\section{Summary and outlook}
The fact that for the Wilson-Dirac operator hermitian conjugation 
can be implemented as a similarity transformation with $\gamma_5$ 
has an important implication for the chiral properties of the 
eigenvectors 
of the fermion matrix: Only eigenvectors with real eigenvalues have
non-vanishing pseudoscalar matrix-elements and thus should be 
interpreted as 
the lattice equivalents of the continuum zero-modes. The lattice
version of the Atiyah-Singer index theorem thus 
connects the topological charge of the spectrum to the number 
of real eigenvalues.

In this contribution we analyzed those features of the spectrum of the
Wilson-Dirac operator that are connected to non-trivial topology of
the gauge field. In particular we demonstrated the stability of the 
real eigenvalues and their corresponding pseudoscalar matrix elements 
under adding fluctuations to the gauge fields. This shows the
topological nature of the real part of the spectrum. 
For the computationally less demanding case of U(1) gauge fields in 
two dimensions we analyzed the lattice index theorem
for a complete set of configurations from a simulation with dynamical 
fermions. It was shown, that the $\beta \rightarrow \infty$ limit is 
dominated by configurations where the lattice
index theorem holds with probability one. 

The next logical step in a numerical 
analysis of the lattice index theorem would
be to test the four-dimensional Wilson-Dirac 
operator with gauge field configurations from a Monte Carlo ensemble 
generated in a quenched or even unquenched 
simulation. On the analytic side it would be interesting to 
investigate the role
of the topological part of the spectrum in fermionic correlation functions. 
In particular in a spectral decomposition of fer\-mio\-nic $n$-point 
functions
one could try to separate the part corresponding to eigenvectors with real 
eigenvalues. The lattice index theorem (\ref{lasit})
shows that these contributions 
are intimately connected to the topological sector of the background gauge 
field. Interesting insights into U(1) problem and Witten-Veneziano type 
formulas could be obtained.
\\
\\
{\bf Acknowledgment:} We thank Christian B.~Lang for valuable discussions
during the course of this work. 
\\
\\
{\bf Note added in the revised version:} In the meantime five more 
contributions related to our work appeared as preprints. In \cite{EdHeNa98}
further studies of the spectral flow of the overlap hamiltonian are 
performed. 
In particular instanton configurations and gauge fields from unquenched 
simulations are considered. Furthermore the effect of an $O(a)$-improving 
clover 
term is discussed. \cite{SiSm98} also analyzes the effect of the clover 
term on the level crossings for the overlap hamiltonian. Improvement seems 
to be
a viable method to enhance the chiral properties of the eigenstates.
In \cite{He98} the spectrum of the un-modified Wilson-Dirac operator is 
analyzed in the Schwinger model and perturbative arguments for the 
stability
of the real eigenvalues are given. Finally in \cite{FaLa98} the spectrum 
of
the fixed point lattice Dirac operator in the Schwinger model is studied.

\end{document}